\documentclass[pre,twocolumn]{revtex4-1}
\usepackage[dvipdfmx]{graphicx}
\usepackage{type1cm}
\usepackage[usenames,svgnames]{xcolor}
\usepackage{color}
\usepackage{url}
\usepackage{natbib}
\usepackage[dvipdfmx]{hyperref}
\hypersetup{
     colorlinks=true,       		
     linkcolor=Navy,          	
     citecolor=Navy,            
     filecolor=Navy,      		
     urlcolor=Navy,           	
    runcolor=cyan,
 }

\usepackage{amsmath,amssymb,bm,amsthm}

\def\beq{\begin{equation}}
\def\eeq{\end{equation}}
\def\nbeq{\begin{equation*}}
\def\neeq{\end{equation*}}

\def\<{\langle}
\def\>{\rangle}

\renewcommand{\d}{\partial}

\begin{document}
\title{Classical ergodicity and quantum eigenstate thermalization: Analysis in fully connected Ising ferromagnets}

\author{Takashi Mori}
\affiliation{
Department of Physics, Graduate School of Science,
University of Tokyo, Bunkyo-ku, Tokyo 113-0033, Japan
}

\begin{abstract}
We investigate the relation between the classical ergodicity and the quantum eigenstate thermalization in the fully connected Ising ferromagnets.
In the case of spin-1/2, an expectation value of an observable in a single energy eigenstate coincides with the long-time average in the underlying classical dynamics, which is a consequence of the Wentzel-Kramers-Brillouin approximation.
In the case of spin-1, the underlying classical dynamics is not necessarily ergodic.
In that case, it turns out that, in the thermodynamic limit, the statistics of the expectation values of an observable in the energy eigenstates coincides with the statistics of the long-time averages in the underlying classical dynamics starting from random initial states sampled uniformly from the classical phase space.
This feature seems to be a general property in semiclassical systems, and the result presented here is crucial in discussing equilibration, thermalization, and dynamical transitions of such systems.
\end{abstract}
\maketitle


\section{Introduction}
\label{sec:intro}

The study of out-of-equilibrium dynamics in isolated quantum systems has received much attention triggered by experimental advance in ultra-cold atomic systems~\cite{Kinoshita2006,Hofferberth2007,Trotzky2012}.
One of the key problems is to understand the property of the steady state reached after sufficiently long times starting from a certain out-of-equilibrium initial state~\cite{Neumann1929, Berry1977, Srednicki1994, Deutsch1991, Rigol2008, Biroli2010, Cassidy2011, Tasaki2016_typicality, Tasaki1998, Berges2004, Reimann2008, Goldstein2010, Calabrese2011, Sato2012, Caux2013,Kaminishi2015, Reimann2015, Vidmar2016}.
Theoretical studies on this fundamental problem have progressed in recent years due to theoretical development and rearrangement of some old fundamental ideas~\cite{Neumann1929,Berry1977,Srednicki1994,Deutsch1991}.
The eigenstate thermalization hypothesis (ETH) is an important notion~\cite{Rigol2008,Biroli2010,Cassidy2011,Tasaki2016_typicality}.
It insists that $\<\phi_n|\mathcal{O}|\phi_n\>\approx\mathrm{Tr}\, \mathcal{O}\rho_{\mathrm{mc}}$ for any energy eigenstate $\phi_n$ and any local observable $\mathcal{O}$, where $\rho_{\mathrm{mc}}$ is the microcanonical density matrix with the energy identical to the energy eigenvalue $E_n$ of $\phi_n$.

ETH is equivalent to the following statement:
\beq
\lim_{T\rightarrow\infty}\frac{1}{T}\int_0^Tdt\<\psi(t)|\mathcal{O}|\psi(t)\>\approx\mathrm{Tr}\, \mathcal{O}\rho_{\mathrm{mc}}
\label{eq:ergodicity}
\eeq
for any initial state picked up from the microcanonical energy shell, $|\psi(0)\>=\sum_nc_n|\phi_n\>$ with $c_n$ nonzero only for $n$ such that $E_n\in[E,E+\Delta E)$, where $\Delta E$ is the energy width in the microcanonical ensemble.
The temporal fluctuation of $\<\psi(t)|\mathcal{O}|\psi(t)\>$ is typically very small for macroscopic systems~\cite{Rigol2008,Short2011,Tasaki2016_typicality}, so that Eq.~(\ref{eq:ergodicity}) implies $\<\psi(t)|\mathcal{O}|\psi(t)\>\approx\mathrm{Tr}\,\mathcal{O}\rho_{\mathrm{mc}}$ for a sufficiently large typical time $t$, which implies thermalization of the system.

From Eq.~(\ref{eq:ergodicity}), ETH is regarded as a quantum counterpart of ergodicity in classical systems.
Naturally, it is expected that there is a close relation between classical ergodicity and quantum ETH.
Indeed, in his pioneering work, Berry conjectured that the classical ergodicity implies the quantum ETH in the semiclassical regime~\cite{Berry1977}.
A recent numerical study~\cite{Russomanno2015} demonstrated this relation in a periodically-driven system.
When the classical dynamics is not ergodic, the classical phase space is divided into the regular and the chaotic regions.
The property of semiclassical eigenstates in such a case has been also extensively studied~\cite{Percival1973,Davis1981,Heller1984,Tomsovic1993,Ketzmerick2000,Hufnagel2002,Backer2008}.
It is known (but not proved) that each energy eigenstate is classified into the regular or chaotic one, corresponding to the regular and the chaotic regions in the classical phase space~\cite{Percival1973}.

The structure of semiclassical energy eigenstates have been mainly studied in the context of quantum chaos. 
In recent years, it has been also investigated in the context of thermalization in isolated quantum systems~\cite{Santos2010,Borgonovi_review2016}.
An apparent connection between the classical ergodicity and the quantum ETH motivates us to further investigate the implication of the quantum-classical correspondence to the problem of thermalization or equilibration in isolated quantum systems.

It should be pointed out that the semiclassical limit of a system with a few degrees of freedom emerges as the thermodynamic limit of a certain kind of quantum many-body systems in the totally symmetric subspace (TSS)~\cite{Sciolla2010,Sciolla2011}, and the result on the quantum-classical correspondence is crucial for the thermalization property of such a system.
A fully connected Ising (anti-)ferromagnet is one such system, and fully connected Ising (anti-)ferromagnets are available in ion-trap experiment~\cite{Porras2004,Britton2012,Islam2013,Richerme2014}.
Therefore, considering the TSS of a quantum many spin system provides a good way to construct a semiclassical system, and is suited to investigate the relation between the quantum ETH and the classical ergodicity.
In this paper, we study the relation between the long-time behavior of the classical dynamics and the property of energy eigenstates in the spin-1/2 and the spin-1 fully connected Ising ferromagnets, the latter of which can show nonergodic classical dynamics in the classical limit.

The organization of this paper is as follows.
In Sec.~\ref{sec:setup}, the setting is explained.
In Sec.~\ref{sec:spin-1/2}, we consider the spin-1/2 case.
We will see in Sec.~\ref{sec:ergodicity} that the ETH within the TSS is equivalent to the classical ergodicity, which is a consequence of the Wentzel-Kramers-Brillouin (WKB) approximation.
The underlying classical dynamics is investigated in Sec.~\ref{sec:trajectory}, and it is demonstrated that the spin-1/2 system thermalizes as a result of the ETH within the TSS in Sec.~\ref{sec:thermalization}.
In Sec.~\ref{sec:dynamical_1/2}, we discuss dynamical transitions, which have been found in several fully connected models~\cite{Sciolla2010,Sciolla2011} and conclude that a dynamical transition is regarded as an equilibrium phase transition within the TSS in the spin-1/2 case.
In Sec.~\ref{sec:spin-1}, we study the spin-1 case.
In the spin-1 case, the WKB approximation is not valid and the ergodicity of the classical dynamics is not ensured.
In Sec.~\ref{sec:statistics}, we will conjecture and numerically verify that the statistics of energy eigenstate expectation values of a local quantity coincides with the statistics of long-time averages of the same quantity in the underlying classical dynamics starting from randomly sampled initial states.
This result is expected to be true for any semiclassical system.
This conjecture implies that the ETH within the TSS holds in the energy region where the underlying classical dynamics is ergodic, while it does not hold in the energy region where the classical dynamics is not ergodic.
Thermalization in the classically ergodic region and no thermalization in the classically nonergodic region are demonstrated in Sec.~\ref{sec:dynamics}.
In Sec.~\ref{sec:relaxation_time}, the system-size dependence of the relaxation time is discussed.
We also study dynamical transitions in the spin-1 model in Sec.~\ref{sec:DPT_spin1}, and in contrast to the spin-1/2 case, it is found that a dynamical transition cannot be necessarily interpreted as an equilibrium phase transition within the TSS because of the lack of ETH in the TSS.
Summary and discussion is given in Sec.~\ref{sec:summary}.

\section{Fully-connected Ising ferromagnets}
\label{sec:setup}

\subsection{Model}
\label{sec:model}

The Hamiltonian is given by
\beq
H=-\frac{J}{2N}\sum_{i\neq j}^N S_i^zS_j^z-h_x\sum_{i=1}^NS_i^x-h_z\sum_{i=1}^NS_i^z+D\sum_{i=1}^N(S_i^z)^2,
\label{eq:Ham}
\eeq
where $\bm{S}_i$ is the spin-1/2 or spin-1 operator of $i$th spin; $J$ is the exchange interaction; $h_x$ and $h_z$ are the magnetic field along $x$ and $z$ directions, respectively; and $D$ is the anisotropic term, which plays the role only for the spin-1 case.
Without $D$ term, this model is known as the Lipkin-Meshkov-Glick (LMG) model~\cite{Lipkin1965}.
We consider the ferromagnetic coupling and put $J=1$, but essentially the same result also holds for antiferromagnetic coupling $J=-1$.
We set $\hbar=1$ throughout the paper.

In fully connected spin systems, there is the permutation symmetry of any $i$th and $j$th spins.
Therefore, if the initial state is totally symmetric, i.e., symmetric for any permutation of spins, the state remains in the TSS during the quantum dynamics.
We assume it, and always consider it in the TSS.

It can be shown that $1/N$ plays the role of the Planck constant in the TSS, and the thermodynamic limit $N\rightarrow\infty$ corresponds to the classical limit~\cite{Sciolla2010,Sciolla2011}.
When $N$ is large but finite, the system described by Eq.~(\ref{eq:Ham}) is regarded as a semiclassical system, and thus these models are good starting points to investigate the relation between the long-time classical dynamics and the property of the quantum eigenstates.
Moreover, fully connected Ising (anti-)ferromagnets can be realized in ion-trap experiment~\cite{Porras2004,Britton2012,Islam2013,Richerme2014}, and we can compare theoretical consequence to experiment.

\subsection{Equilibration after a quench}
\label{sec:quench}

We will consider the quantum dynamics after a quench, i.e., a sudden change of parameters of the Hamiltonian.
The initial state at time $t=0$ is chosen as the ground state of the prequench Hamiltonian, which is in the TSS, and hence the quantum state at time $t>0$ is always in the TSS.
Since the pre-quench Hamiltonian is the quantum many-body Hamiltonian and the energy per spin is independent of $N$, the fluctuation of the energy density $\delta\varepsilon$ after the quench is typically proportional to $N^{-1/2}$.
Therefore, we can consider that the system has an almost definite value of the energy density $\varepsilon$ after a quench.
In other words, we can consider the quantum dynamics within a suitable microcanonical energy shell of the TSS. 

If the initial state $|\psi(0)\>=\sum_nc_n|\phi_n\>$ is given by a superposition of a large number of energy eigenstates, the temporal fluctuation of $\<\psi(t)|\mathcal{O}|\psi(t)\>$ is very small.
It is shown that for the effective dimension $d_{\mathrm{eff}}$ of the initial state defined as $d_{\mathrm{eff}}:=\left(\sum_n|c_n|^4\right)^{-1}$, the temporal fluctuation given by $\left(\overline{\<\psi(t)|\mathcal{O}|\psi(t)\>^2}-\overline{\<\psi(t)|\mathcal{O}|\psi(t)\>}^2\right)^{1/2}$ is smaller than a quantity of $O(d_{\mathrm{eff}}^{-1/2})$ under the non-resonance condition~\cite{Short2011}.
In the case of fully connected Ising ferromagnets, the number of energy eigenstates in the TSS with the energy density between $\varepsilon$ and $\varepsilon+\delta\varepsilon$ is proportional to $N\delta\varepsilon$ in the case of spin-1/2 and $N^2\delta\varepsilon$ in the case of spin-1.
As mentioned above, the fluctuation of the energy density after a quench is given by $\delta\varepsilon\sim N^{-1/2}$, and hence $d_{\mathrm{eff}}$ is typically very large (scaled as $d_{\mathrm{eff}}\sim N^{1/2}$ for spin-1/2 and $N^{3/2}$ for spin-1).
It is therefore expected that $\<\psi(t)|\mathcal{O}|\psi(t)\>$ will reach an almost stationary value without any large fluctuation after a sufficiently long time.

In this way, fully connected ferromagnets \textit{equilibrate} (approach to a stationary state) within the microcanonical energy shell after a quench.
However, it is nontrivial whether the system thermalizes within the TSS.
It depends on the property of semiclassical energy eigenstates, which will be related to the property of classical dynamics.

From now on, we investigate the property of semiclassical energy eigenstates both for spin-1/2 case and for spin-1 case and discuss its consequence in equilibration, thermalization, and dynamical transitions after a quench.

\section{Spin-1/2 case}
\label{sec:spin-1/2}

\subsection{Ergodicity and ETH}
\label{sec:ergodicity}

Let us start  from the simple spin-1/2 case.
The TSS corresponds to the subspace with the maximum total spin,
$\left(\sum_{i=1}^N\bm{S}_i\right)^2=(N/2)(N/2+1)$.
The energy eigenstates are labeled by a single variable $z$, where
\beq
\frac{1}{N}\sum_{i=1}^NS_i^z|z\>=z|z\>.
\eeq
The wave function is defined as $\psi_t(z):=\<z|\psi(t)\>$ for a state $|\psi(t)\>$, and then, it is shown that the Schr\"odinger equation $id|\psi(t)\>/dt=H|\psi(t)\>$ reduces to
\begin{align}
\frac{i}{N}\frac{\d}{\d t}\psi_t(z)&=\left(-\frac{1}{2}z^2-h_zz-h_x\sqrt{\frac{1}{4}-z^2}\cos p_z\right)\psi_t(z)
\nonumber \\
&=:\tilde{H}\psi_t(z)
\label{eq:Sch}
\end{align}
in the leading order in $1/N$~\cite{Sciolla2010,Sciolla2011}.
Here, $1/N$ plays the role of the Planck constant, and the canonical momentum conjugate to $z$ is defined as $p_z:=(-i/N)\d/\d z$.

When $N\gg 1$, an energy eigenstate $\phi_{\varepsilon}(z)=\<z|\phi_n\>$ with an energy eigenvalue $E_n=N\varepsilon$ is given by the WKB approximation.
The immediate consequence of the WKB approximation is that the eigenstate expectation value of an observable $f(z,p_z)$, where $f$ is a function independent of $N$, asymptotically equals the long-time average of the same quantity in the underlying classical dynamics for large $N$~\cite{Sciolla2010,Sciolla2011},
\begin{align}
\<\phi_n|f(z,p_z)|\phi_n\>&\approx\lim_{T\rightarrow\infty}\frac{1}{T}\int_0^Tdtf(z(t),p_z(t))
\nonumber \\
&=:\overline{f(z(t),p_z(t))},
\label{eq:average}
\end{align}
where $z(t)$ and $p_z(t)$ are the solutions of the classical equations of motion, $dz(t)/dt=\d\tilde{H}/\d p_z$ and $dp_z(t)/dt=-\d\tilde{H}/\d z$.
Because the equal energy surface $\tilde{H}(z,p_z)=\varepsilon$ is one dimensional in the phase space, the ergodicity of the classical dynamics is trivial as long as the equal energy surface is simply connected.
As a result,
\beq
\<\phi_n|f(z,p_z)|\phi_n\>\approx\overline{f(z(t),p_z(t))}\approx f_{\mathrm{eq}},
\eeq
where $f_{\mathrm{eq}}$ is the equilibrium value of $f$ calculated in the microcanonical ensemble of $H$ with the energy $N\varepsilon$ within the TSS~\footnote{$f_{\mathrm{eq}}$ is not equal to the expectation value in the microcanonical ensemble of the whole Hilbert space. We should consider the equilibrium state within the TSS.}.
This implies that quantum energy eigenstates $\phi_n$ satisfy ETH within the TSS~\footnote{It might be better to call the ETH within the totally symmetric subspace the ``generalized eigenstate thermalization'' in order to distinguish it from the ETH in the whole Hilbert space, see Ref.~\cite{Cassidy2011,Vidmar2016}}.

In this model, in some choice of parameters $\{h_x,h_z,\varepsilon\}$, the equal-energy surface of $\tilde{H}$ is separated into the two disconnected parts (ergodic regions), see Fig.~\ref{fig:equal-energy}.
In that case, the energy eigenstates are also divided into the two sectors, corresponding to the two separated ergodic regions of the underlying classical dynamics, and the ETH holds for each sector although the ETH does not hold as a whole.
In the spin-1/2 case, this correspondence between the long-time average in the classical dynamics and the quantum mechanical average in a single energy eigenstate is a consequence of the WKB approximation.

\subsection{Classical trajectories and the eigenstate expectation values}
\label{sec:trajectory}

\begin{figure}[tb]
\begin{center}
  \includegraphics[width=70mm]{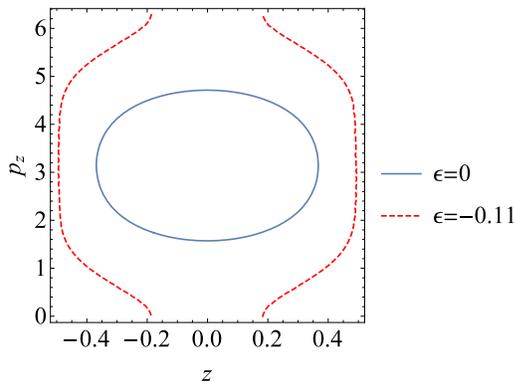}
\end{center}
 \caption{The equal energy surfaces for $\varepsilon=0$ (solid line) and for $\varepsilon=-0.11$ (dashed line). The parameters are chosen as $h_x=0.2$ and $h_z=0.001$.}
 \label{fig:equal-energy}
\end{figure}

\begin{figure}[tb]
\begin{center}
  \includegraphics[width=70mm]{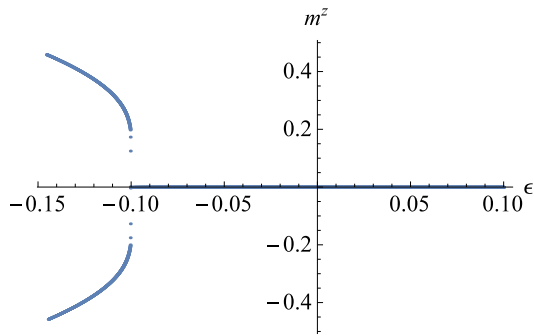}
\end{center}
 \caption{The eigenstate expectation values $\<\phi_n|m^z|\phi_n\>$.
The horizontal axis is the energy eigenvalue divided by $N$.
The parameters are chosen as $h_x=0.2$, $h_z=0.001$, and $N=5000$.}
 \label{fig:eigenstate}
\end{figure}

The classical trajectory is fully determined by the equal-energy surface $\tilde{H}(z,p_z)=\varepsilon$ because it is one dimensional in the phase space.
Typical shapes of the equal-energy surfaces are given in Fig.~\ref{fig:equal-energy}.
The parameters are chosen as $h_x=0.2$ and $h_z=0.001$.
The equal energy surface $\tilde{H}(z,p_z)=\varepsilon$ with $\varepsilon=0$ is depicted by the solid line and that with $\varepsilon=-0.11$ is depicted as the dashed line.
For $\varepsilon=0$, the equal-energy surface is simply connected.
If any point on the equal-energy surface is chosen as an initial state, the classical trajectory passes through all the points on the same equal-energy surface, i.e. the classical dynamics is trivially ergodic.
On the other hand, for $\varepsilon=-0.11$, the equal-energy surface is divided into the two disconnected regions (it is noted that $p_z=0$ is equivalent to $p_z=2\pi$).
In that case, the classical orbit is given by either of the two curves depending on the initial state, and the corresponding quantum energy eigenstates are also divided into the two branches, shown in Fig.~\ref{fig:eigenstate}.

It should be pointed out that $h_z=0$ is an exception.
When $h_z=0$, two disconnected equal-energy surfaces with $\overline{m^z(t)}>0$ and $\overline{m^z(t)}<0$ appear in the classical dynamics, but the quantum expectation values of $m^z$ are zero, $\<\phi_n|m^z|\phi_n\>=0$ for all $n$ because of the inversion symmetry of the $z$ component of the magnetization.
It is interpreted that the two equal-energy surfaces are connected through the resonant quantum tunneling.
However, as long as $h_z$ is nonzero and there is no symmetry between the two disjoint ergodic regions, such resonant tunneling is suppressed and we can see that $\<\phi_n|m^z|\phi_n\>>0$ for some $n$ and $\<\phi_n|m^z|\phi_n\><0$ for the others when $N$ is large.

\subsection{Thermalization within the TSS}
\label{sec:thermalization}

We shall investigate the quantum dynamics after a quench in this model.
Here, the parameters $\{h_x,h_z\}$ are suddenly quenched from $\{h_x^{(i)},h_z^{(i)}=0.1\}$ to $\{h_x^{(f)}=0.2,h_z^{(f)}=0.001\}$, and the other parameters are fixed to be the same values as in the previous section.
Then, the post-quench Hamiltonian is identical to the Hamiltonian considered in the previous section.

Note that the initial value of $h_z^{(i)}=0.1$ will choose the sector of $m^z>0$ in the region of $\varepsilon\lesssim -0.1$, in which energy eigenstates are divided into the two sectors.
Therefore, we can neglect the presence of the sector of $m^z<0$, and hence the ETH practically holds for the entire energy region.

\begin{figure}[tb]
\begin{center}
\begin{tabular}{c}
\includegraphics[width=70mm]{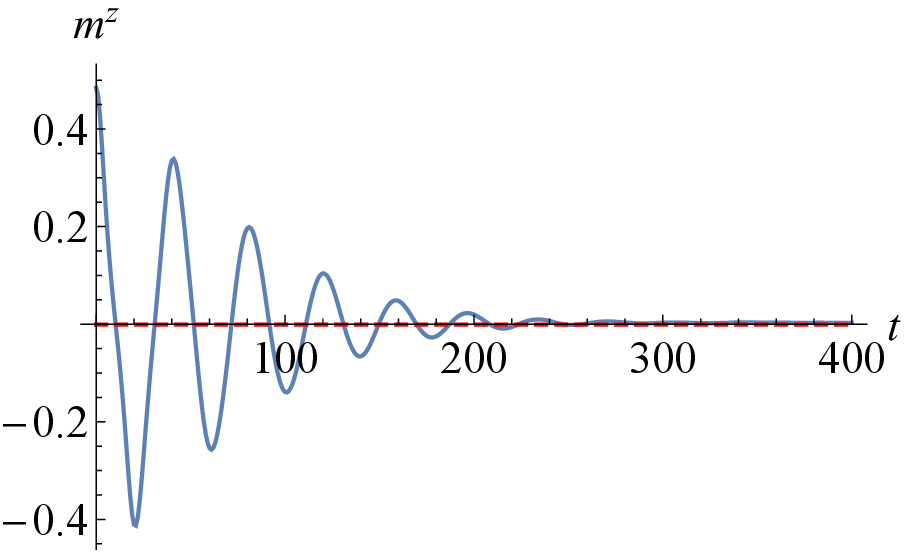}\\
\includegraphics[width=70mm]{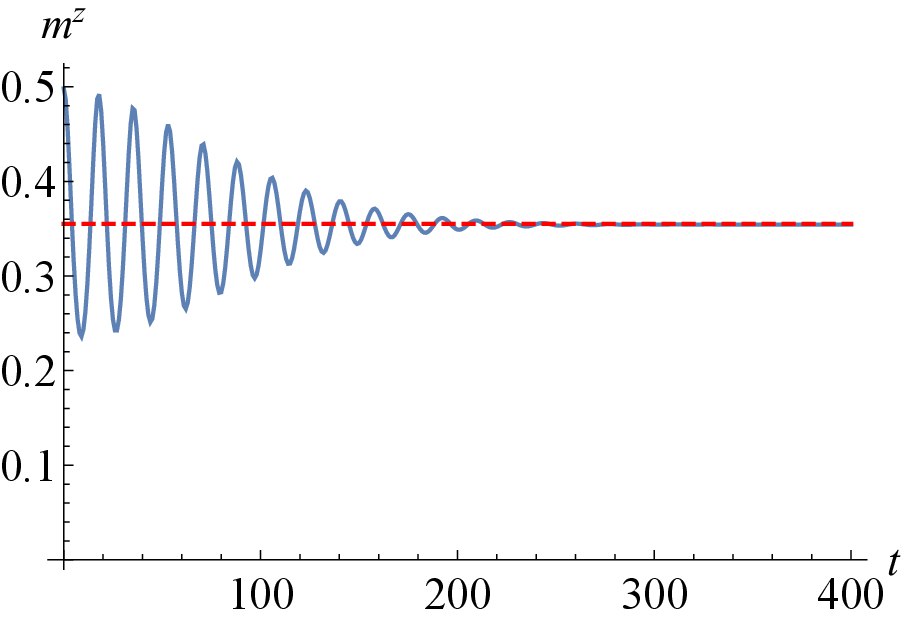}
\end{tabular}
\end{center}
\caption{Time evolutions of $\<\Psi(t)|m^z|\Psi(t)\>$ after the quenches from (top) $h_x^{(i)}=-0.15$ and (bottom) $h_x^{(i)}=-0.05$.
The solid lines are the solutions of the Schr\"odinger equation for $N=4000$ and the dashed red lines are the equilibrium values in the TSS.}
\label{fig:quench_Lipkin}
\end{figure}

In Fig.~\ref{fig:quench_Lipkin}, time evolutions of $m^z:=(1/N)\sum_{i=1}^NS_i^z$ are shown for $h_x^{(i)}=-0.15$ (top) and $-0.05$ (bottom).
Red dashed lines represent the equilibrium value within the TSS.
It is clear that the system reaches equilibrium within the TSS after long times, which results from the ETH within the TSS.

The timescale in which observables reach their stationary values is numerically found to be proportional to $N^{1/2}$.
In the limit of $N\rightarrow\infty$, the classical dynamics is exact forever, and the relaxation to the steady state does not occur.
This size dependence of the relaxation time is consistent with the result on the exactly solvable Emch-Radin model (no $h_x$ and $D$ terms)~\cite{Kastner2011,Worm2013}, and is also consistent with the semiclassical evaluation, see Sec.~\ref{sec:relaxation_time}.

\subsection{Dynamical phase transitions after a quench}
\label{sec:dynamical_1/2}

\begin{figure}[tb]
\begin{center}
  \includegraphics[width=70mm]{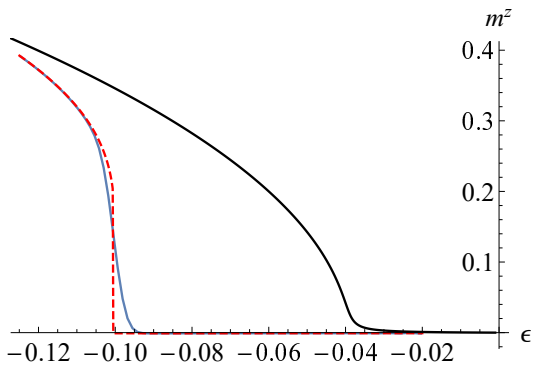}
\end{center}
 \caption{Blue (gray) solid line: Expectation values of $m^z$ in the steady state after quenches as a function of $\varepsilon$ ($\varepsilon$ is a function of $h_x^{(i)}$).
Red dashed line: Expectation values of $m^z$ in the equilibrium state within the TSS, which is the same one as the curve in Fig.~\ref{fig:eigenstate}.
Black solid line: Expectation values of $m^z$ in the true equilibrium state (not restricted to the TSS).}
 \label{fig:DPT}
\end{figure}

In Fig.~\ref{fig:DPT}, we show the expectation values of $m^z$ (blue solid line) in the steady state reached after quenches for various values of $h_x^{(i)}$.
The transverse axis is the energy density after the quench, which is a function of $h_x^{(i)}$.
The data agree very well with the equilibrium values within the TSS (red dotted line, which is the same one as the curve of Fig.~\ref{fig:eigenstate}), which shows the phase transition around at $\varepsilon=-0.1$.
This transition is called a \textit{dynamical transition} because its transition point differs from the transition point in the \textit{true thermal equilibrium} (black solid line in Fig.~\ref{fig:DPT}), i.e., the equilibrium state not restricted to the TSS.

Dynamical phase transitions in several fully connected systems have been studied by Sciolla and Biroli~\cite{Sciolla2010,Sciolla2011}.
Figure~\ref{fig:DPT} shows that a dynamical transition is nothing but an equilibrium phase transition within the TSS \textit{when the underlying classical Hamiltonian has only one degree of freedom}.

\section{Spin-1 case}
\label{sec:spin-1}

\subsection{Statistics of classical long-time averages and that of energy eigenstate expectation values}
\label{sec:statistics}

\begin{figure*}[tb]
\begin{center}
\begin{tabular}{ccc}
\includegraphics[width=55mm]{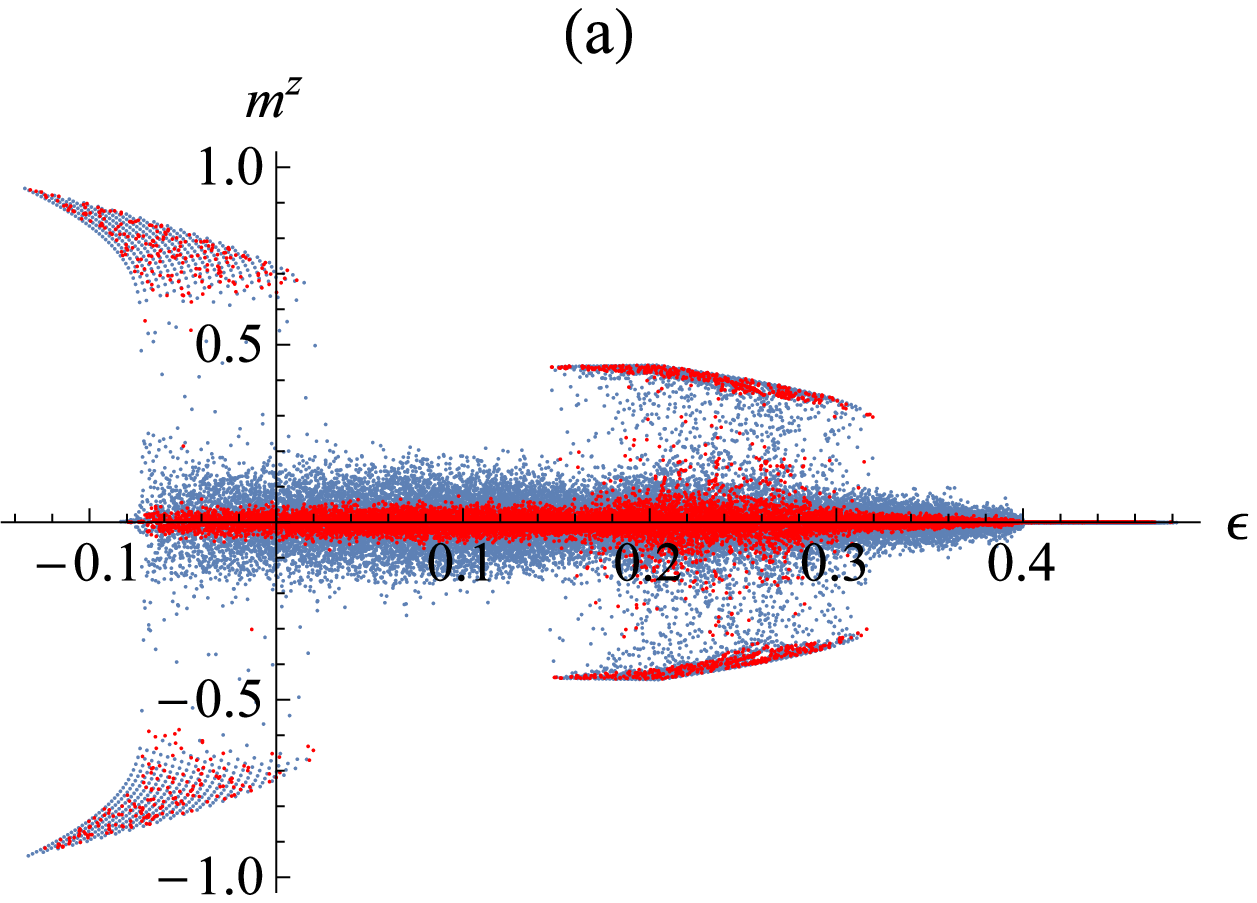}&
\includegraphics[width=55mm]{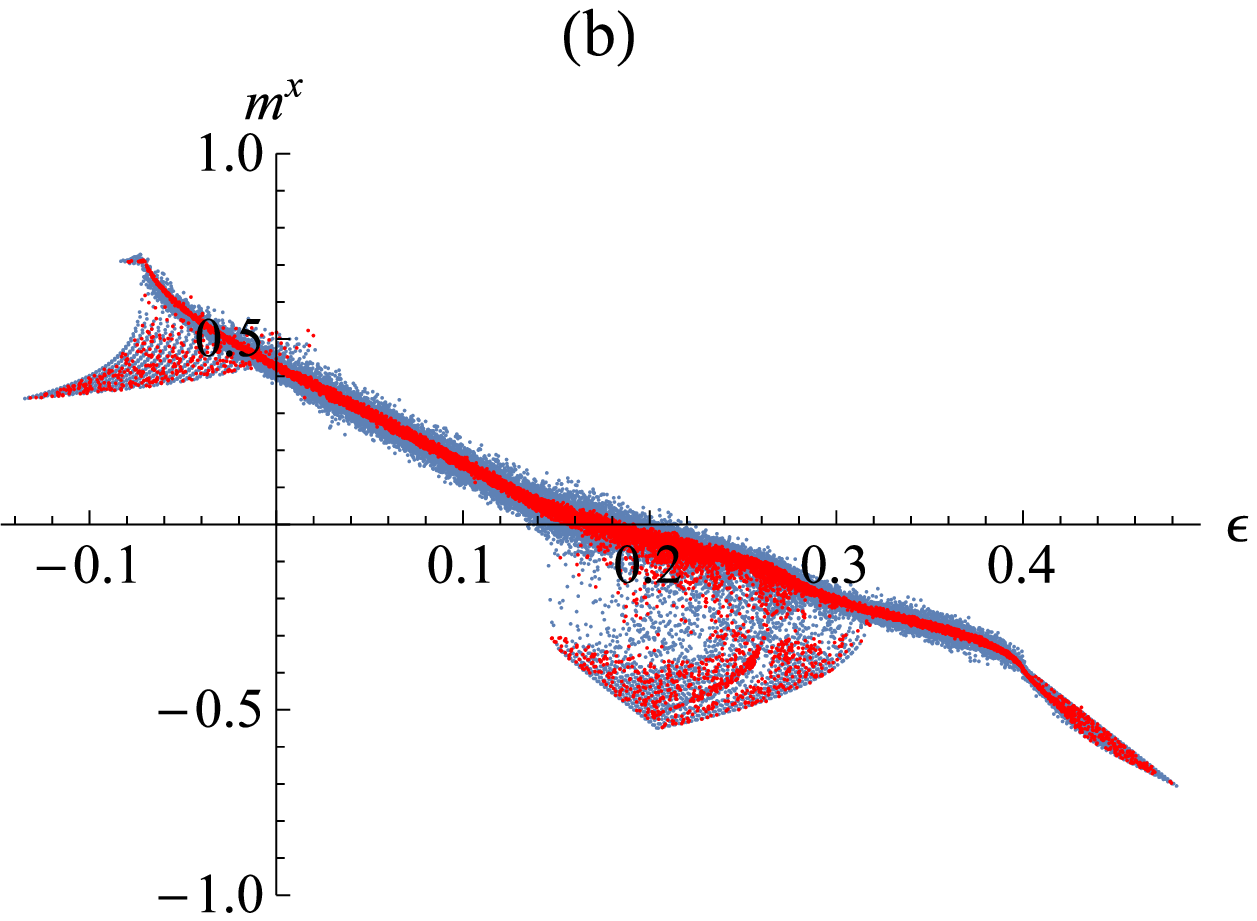}&
\includegraphics[width=55mm]{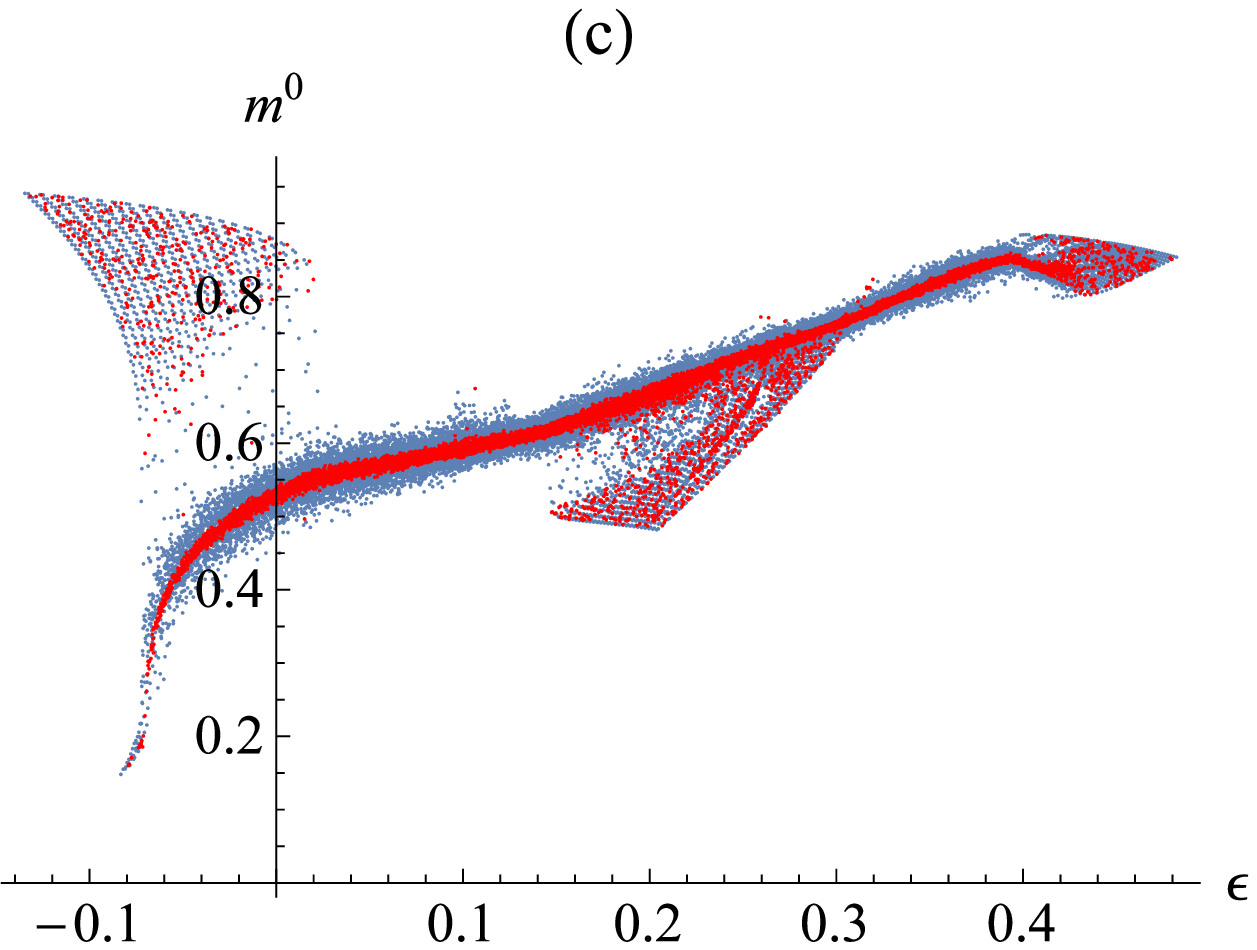}
\end{tabular}
\end{center}
\caption{Long-time averages of (a) $m^z$, (b) $m^x$, and (c) $m^0$ in the classical dynamics starting from random initial states sampled uniformly from the phase space (red points) and expectation values of the same quantities in each quantum energy eigenstates for $N=240$ (blue points). The horizontal axis is the energy density $\varepsilon$.
The parameters are set as $h_x=0.2$, $h_z=0.01$, and $D=0.4$.
The number of the samples of the random initial states in the classical dynamics is 14000.
The classical dynamics is calculated up to $t=20000$ and the long-time averages are calculated within this time interval.}
\label{fig:classical-quantum}
\end{figure*}

In the spin-1 case, the situation becomes complicated.
Because of the presence of the anisotropy $D\sum_{i=1}^N(S_i^z)^2$, the total spin is not conserved.
The energy eigenstates in the TSS are characterized by the two variables $x$ and $y$, where
\beq
\left\{
\begin{split}
\frac{1}{N}\sum_{i=1}^N\frac{S_i^z(S_i^z+1)}{2}|x,y\>=x|x,y\>, \\
\frac{1}{N}\sum_{i=1}^N\frac{S_i^z(S_i^z-1)}{2}|x,y\>=y|x,y\>.
\end{split}
\right.
\label{eq:xy}
\eeq
The number of the spins in $S_i^z=+1$ $(-1)$ equals $Nx$ ($Ny$).
Similarly to the spin-1/2 case, by deriving the Schr\"odinger equation for the wave function $\psi_t(x,y):=\<x,y|\psi(t)\>$ up to the leading order in $1/N$, we obtain
\beq
\frac{i}{N}\frac{\d}{\d t}\psi_t(x,y)=\bar{H}\psi_t(x,y),
\eeq
with
\begin{align}
\bar{H}(x,y,p_x,p_y)=-\frac{1}{2}(x-y)^2+D(x+y)-h_z(x-y)
\nonumber \\
-h_x\left(\sqrt{2x(1-x-y)}\cos p_x +\sqrt{2y(1-x-y)}\cos p_y\right).
\end{align}
Again the effective Planck constant is given by $1/N$ and the canonical momenta are defined as $p_x=(-i/N)\d/\d x$ and $p_y=(-i/N)\d/\d y$.
The underlying classical dynamics $\{x(t),y(t),p_x(t),p_y(t)\}$ is given by the Hamilton equation in terms of the classical Hamiltonian $\bar{H}$, but now there are two degrees of freedoms ($x$ and $y$).
The equal energy surface $\bar{H}=\varepsilon$ is three dimensional, and the classical ergodicity is not trivial.
In Appendix~\ref{sec:Poincare}, some Poincare sections are presented.

As in the spin-1/2 case, it is expected that there is a close relation between the long-time average in the classical dynamics and the quantum energy eigenstate expectation value in the spin-1 case, although we cannot apply the WKB approximation.

\begin{figure}[tb]
 \begin{center}
  \includegraphics[width=70mm]{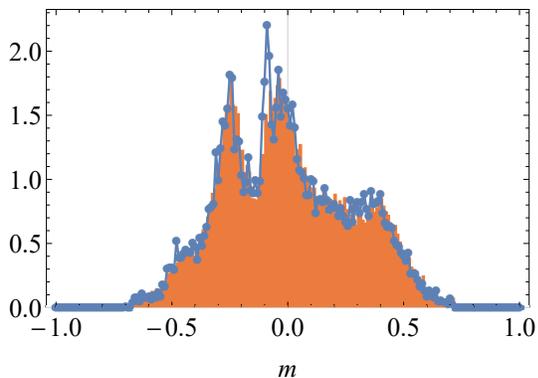}
\end{center}
 \caption{$P_N(m)$ for $N=240$ (shaded histogram) and $P_{\mathrm{cl}}(m)$ (points and line).}
 \label{fig:distribution}
\end{figure}

Here, we shall compare the distribution of the long-time averages of some quantities in the classical dynamics with the initial states sampled randomly from the phase space to the distribution of the energy eigenstate expectation values of the same quantities.
In Fig.~\ref{fig:classical-quantum}, the results for (a) $m^z=(1/N)\sum_{i=1}^NS_i^z=x-y$, (b) $m^x=(1/N)\sum_{i=1}^NS_i^x\approx\sqrt{2x(1-x-y)}\cos p_x+\sqrt{2y(1-x-y)}\cos p_y$, and (c) $m^0=(1/N)\sum_{i=1}^N[1-(S_i^z)^2]=1-x-y$ are shown.
The transverse axis is the energy density $\varepsilon$.
The red (blue) points are the long-time averages in the classical dynamics $\overline{\mathcal{O}(t)}$ (the energy eigenstate expectation values $\<\phi_n|\mathcal{O}|\phi_n\>$) for $\mathcal{O}=m^z, m^x,$ and $m^0$.
They agree very well including the strongly nonergodic energy region.
The distribution functions also agree very well.
In Fig.~\ref{fig:distribution}, $P_N(m)$ and $P_{\mathrm{cl}}(m)$ are shown for $N=240$, where $P_N(m)$ is the probability distribution function of $m=\<\phi_n|m^x|\phi_n\>$ obtained by all the energy eigenstates in a finite $N$ quantum system, and $P_{\mathrm{cl}}(m)$ is the probability distribution function of the long-time average of $m^x(t)$ in the underlying classical dynamics starting from randomly sampled initial states.
They agree very well.
We can also consider the probability distribution function of $m^z$ or $m^0$, but the result does not change.

\begin{figure}[tb]
 \begin{center}
  \includegraphics[width=70mm]{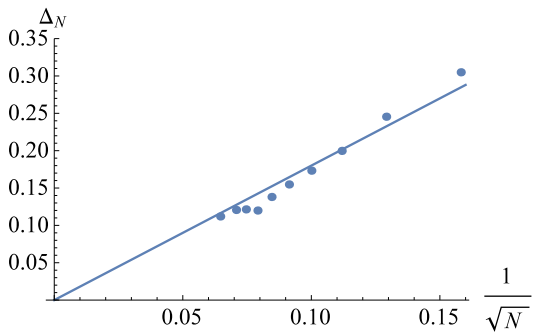}
\end{center}
 \caption{$\Delta_N$ as a function of $1/\sqrt{N}$. The smallest value and the largest value of $N$ are 40 and 240, respectively.}
 \label{fig:finite-size}
\end{figure}

Next, we consider the finite-size scaling of the deviation of the two distributions defined as
\beq
\Delta_N:=\int dm |P_N(m)-P_{\mathrm{cl}}(m)|.
\eeq
In Fig.~\ref{fig:finite-size}, we see that $\Delta_N$ decreases towards zero as $\Delta_N\propto N^{-1/2}$ at least up to $N=240$.
This finite-size analysis strongly indicates that the two distributions coincide in the limit of $N\rightarrow\infty$.

Although we only show the result for the fully connected Ising ferromangets, the similar result will hold for other semiclassical systems.
Indeed, in the Dicke model, which is regarded as another semiclassical model, the same result is confirmed, see Appendix~\ref{sec:Dicke}.
From these observations in the fully connected models, \textit{it is conjectured that this coincidence of the distribution of the quantum eigenstate expectation values and that of the long-time averages in the underlying classical dynamics is a general feature in semiclassical systems}.

The above conjecture is consistent with the previous studies.
As mentioned earlier, it is known (but not proved) that the energy eigenstates are classified into regular or irregular ones~\cite{Percival1973} correspondingly to the regular and the irregular (chaotic) regions in the classical phase space~\footnote{For small but finite values of the Planck constant, there are energy eigenstates localized neither on integrable nor the whole chaotic regions~\cite{Ketzmerick2000}.}.
Intuitively, regular eigenstates will behave as ones obtained by the WKB theory, and the connection between the energy eigenstate average and the classical infinite time average is understood as the result of the WKB theory, as discussed in the spin-1/2 case.
For irregular eigenstates, the coincidence of the quantum and classical averages can be viewed as a variant of Berry's conjecture on ergodic systems.

\subsection{Steady state after a quench}
\label{sec:dynamics}

Let us discuss the quantum dynamics of the spin-1 fully connected Ising ferromagnet based on our conjecture on the semiclassical limit. 
The conjecture implies that the quantum ETH is satisfied within the TSS if the underlying classical dynamics is ergodic.
On the other hand, if the classical dynamics is not ergodic, then the quantum ETH is violated.
In that case, according to our conjecture, the statistics of $\<\phi_n|\mathcal{O}|\phi_n\>$ for many different $n$ such that $E_n/N\in[\varepsilon,\varepsilon']$ for arbitrary $\varepsilon<\varepsilon'$ coincides with the statistics of $\overline{\mathcal{O}(t)}$ for random initial states sampled uniformly from the phase space with the energy between $\varepsilon$ and $\varepsilon'$.

As argued in Sec.~\ref{sec:quench}, the system equilibrates after a quench and the energy density after a quench is given by $\varepsilon$ with a very small fluctuation $\delta\varepsilon$.
Our conjecture implies that the stationary value of the expectation value of an observable will be equal to the equilibrium value within the TSS when the classical dynamics is ergodic, but this is not the case when the classical dynamics is not ergodic.

\begin{figure}[tb]
 \begin{center}
\begin{tabular}{c}
\includegraphics[width=70mm]{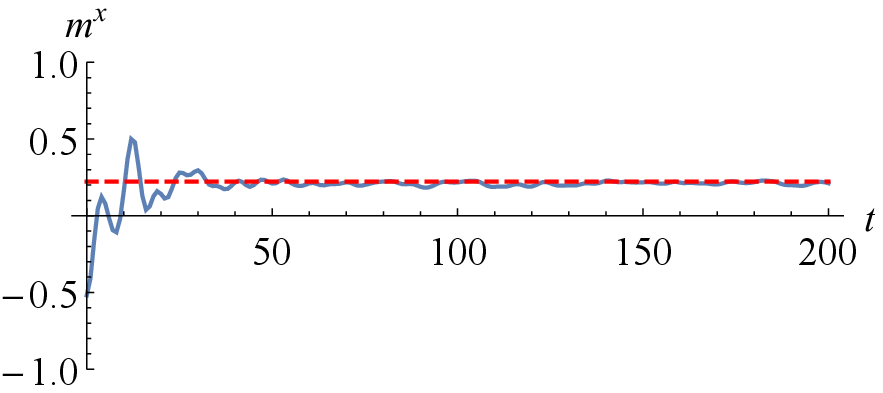}\\
\includegraphics[width=70mm]{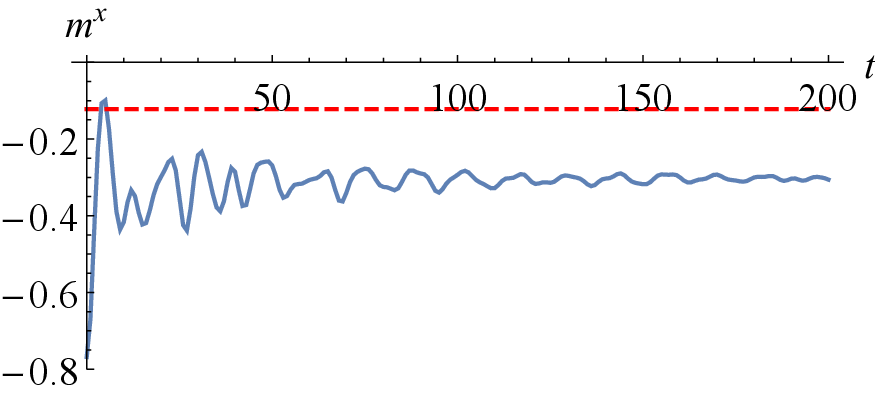}
\end{tabular}
\end{center}
 \caption{Time evolutions of $\<\psi(t)|m^x|\psi(t)\>$ for the quenches of (top) $(h_x^{(i)}=-0.3, h_x^{(f)}=0.2)$ and (bottom) $(h_x^{(i)}=-0.39, h_x^{(f)}=0.2)$.
The solid lines are the solutions of the Schr\"odinger equation for $N=80$.
The dashed red lines are the equilibrium values.}
 \label{fig:dynamics}
\end{figure}

In order to check the above scenario on quantum dynamics for spin-1 case, we numerically calculate the dynamics of $\<m^x(t)\>:=\<\psi(t)|m^x|\psi(t)\>$ after the quench, in which $h_x$ is suddenly quenched from $h_x^{(i)}$ to $h_x^{(f)}=0.2$.
Other parameters are fixed as $h_z=0.01$ and $D=0.4$, and the postquench Hamiltonian is the same as the Hamiltonian employed in Fig.~\ref{fig:classical-quantum}.
The initial state is given as the ground state of the prequench Hamiltonian.

In the top of Fig.~\ref{fig:dynamics}, the time evolution in the case of $h_x^{(i)}=-0.3$ is shown.
In this case, the energy density after the quench is 0.07, in which the classical dynamics is ergodic, see Fig.~\ref{fig:classical-quantum}.
Clearly, $\<m^x(t)\>$ approaches the stationary value almost identical to the equilibrium value within the TSS (the red dashed line).

In the bottom of Fig.~\ref{fig:dynamics}, the dynamics in the quench from $h_x^{(i)}=-0.39$ is shown.
The energy density after the quench is 0.21, in which the classical dynamics is strongly nonergodic, see Fig.~\ref{fig:classical-quantum}.
In this case, as is clearly observed in the bottom of Fig.~\ref{fig:dynamics}, $\<m^x(t)\>$ approaches a stationary value but it differs from the equilibrium value within the TSS.
The absence of thermalization within the TSS is a consequence of the lack of ETH within the TSS in the classically nonergodic region.

\subsection{Relaxation time}
\label{sec:relaxation_time}

In the leading order in the (effective) Planck constant $\hbar_{\mathrm{eff}}=1/N$, it is known that the truncated Wigner approximation is valid~\cite{Walls_text,Polkovnikov2010}.
In this approximation, the initial state is represented as a (quasi-)probability distribution in the classical phase space (Wigner function), and it obeys the classical equations of motion, i.e., the Liouville equation.
Since the intensive quantities $x$, $y$, $p_x$, and $p_y$ show fluctuations proportional to $N^{-1/2}$ in the initial state after a quench, this initial state is represented as a sharply localized distribution function in the classical phase space.
The classical time evolution results in the spread of the distribution over the equal energy surface, and the distribution function will reach some stationary distribution after a ``mixing time''~\footnote
{It is noted that the phase space volume is conserved owing to the Liouville theorem, and therefore the classical distribution function ``spreads'' and ``approach the stationary distribution'' only in a weak sense after a coarse graining of the phase space.
}.
In this picture, the relaxation time of a semiclassical system is given by the mixing time in the underlying classical dynamics.

Suppose that a region of small volume $\Delta V(0)$ in the classical phase space evolves under the classical equations of motion.
According to Krylov~\cite{Krylov_text}, this region will be spread over a region with the volume $\Delta V(t)=\Delta V(0)e^{h_{\mathrm{KS}}t}$ after a time $t$~\cite{Dellago1997}, where $h_{\mathrm{KS}}>0$ is the Kolmogorov-Sinai entropy.
This exponential growth is understood by the fact that the distance between two points close to each other in the phase space grows exponentially fast in the chaotic dynamics.
In the initial state after a quench in our spin-1 model, $\Delta V(0)\sim 1/N^2$, and hence, the mixing time $\tau$ is evaluated as $\tau\propto\ln N$.

On the other hand, when the classical dynamics is regular, the distance between two points in the phase space grows only linearly in $t$.
Since the width of the initial distribution is proportional to $N^{-1/2}$, the mixing time $\tau$ is evaluated as $\tau\propto N^{1/2}$, which is consistent with the relaxation time in the spin 1/2 case discussed in Sec.~\ref{sec:thermalization}.

As shown in Appendix~\ref{sec:Poincare}, the spin-1 model shows regular, chaotic, and mixed (partly regular and partly chaotic) dynamics depending on the energy.
The above discussion indicates that the relaxation time scales as $\ln N$ when the initial quantum state corresponds to the classical phase-space distribution localized in the chaotic region, and as $N^{1/2}$ otherwise.
Although it is hard to distinguish $\ln N$ and $N^{1/2}$ dependence in numerical calculations up to $N=240$, numerical results (not shown) look consistent with those $N$ dependencies of the relaxation times.

\subsection{Dynamical transitions not interpreted as equilibrium phase transitions within the TSS}
\label{sec:DPT_spin1}

\begin{figure}[tb]
\begin{center}
  \includegraphics[width=70mm]{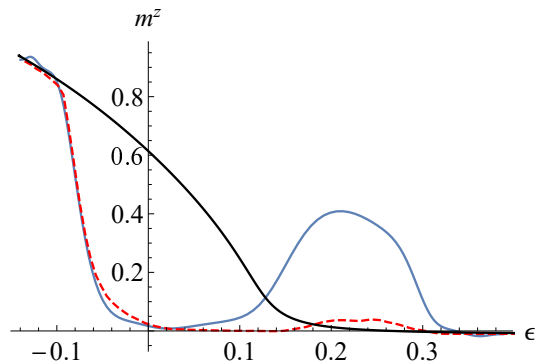}
\end{center}
 \caption{Blue (gray) solid line: Expectation values of $m^z$ in the steady state after quenches as a function of $\varepsilon$, the energy density after the quench.
Red dashed line: Expectation values of $m^z$ in the equilibrium state within the TSS.
Black solid line: Expectation values of $m^z$ in the true equilibrium state (not restricted to the TSS).}
 \label{fig:DPT_spin1}
\end{figure}

Dynamical transitions also take place in the spin-1 model.
Figure~\ref{fig:classical-quantum} (a) shows that energy eigenstates with nonzero magnetizations along the $z$ axis appear.
Since the model with $h_z=0$ has the symmetry of the $\pi$ rotation of every spin along the $x$ axis, these symmetry-breaking energy eigenstates will result in dynamical transitions as in the spin-1/2 case (here we introduced a very small symmetry breaking field $h_z=0.01$ in order to choose the sector of the positive magnetization $m^z>0$). 
The difference from the spin-1/2 case is that symmetry-breaking energy eigenstates appear when the classical dynamics is not ergodic, and hence the ETH does not hold even if we restrict ourselves into the sector of the positive magnetization.
As we saw in Sec.~\ref{sec:dynamics}, the steady state reached after a quench is not fully determined by the energy density after a quench; it also depends on the choice of the prequench Hamiltonian.
As a result, the nature of a dynamical transition, i.e., the location of the transition or the magnetization curve, will depend on the choice of the prequench Hamiltonian.

We consider the same quench protocol as in Sec.~\ref{sec:dynamics}, i.e., we suddenly change the magnetic field along $x$ direction from $h_x^{(i)}$ to $h_x^{(f)}=0.2$, and other parameters are fixed as $h_z=0.01$ and $D=0.4$.
We vary the value of $h_x^{(i)}$, which determines the energy density $\varepsilon$ after the quench.
In Fig.~\ref{fig:DPT_spin1}, we show the numerical result on the expectation value of $m^z$ in the stationary state after a quench as a function of $\varepsilon$ (blue solid line).
It is observed that $m^z$ takes large values in the low-energy regime ($\varepsilon\lesssim 0$) and in the intermediate energy regime ($0.1\lesssim\varepsilon\lesssim 0.3$), which shows dynamical transitions in the stationary state after the quench.

We find that the magnetization curve in the stationary state after the quench differs from the true equilibrium curve (black solid line), which implies that the stationary state is not a true equilibrium state.
In addition, the magnetization curve is also deviated from the equilibrium curve within the TSS (red dashed line), which implies that the stationary state also differs from the equilibrium state within the TSS.
This is due to the absence of the ETH within the TSS.
Thus the dynamical transition in the spin-1 case is not interpreted as an equilibrium phase transition within the TSS in contrast to the spin-1/2 case.

\section{Summary and discussion}
\label{sec:summary}

In summary, the relation between the property of individual energy eigenstates and the long-time behavior of the underlying classical dynamics has been investigated in fully connected Ising ferromagnets, and it has been shown that the distribution of the expectation values in the quantum energy eigenstates converges to the distribution of the long-time averages in the classical dynamics starting from random initial states sampled uniformly from the classical phase space.
This result implies that a fully connected quantum many-body system equilibrates but does not thermalize within the TSS in the classically nonergoric region, as clearly demonstrated in Fig.~\ref{fig:dynamics}.

It should be noted that the nature of an equilibrium state in the TSS strikingly differs from the true equilibrium state defined on the entire Hilbert space.
In the latter, the fluctuation of any macroscopic quantity vanishes in the thermodynamic limit, while in the totally symmetric equilibrium state, the fluctuation of a macroscopic quantity cannot be neglected even in the thermodynamic limit.
In order to understand this aspect, recall that the variables $x$ and $y$ originally represent the density of up and down spins, respectively, see Eq.~(\ref{eq:xy}), which are macroscopic intensive quantities.
In the classical dynamics, $x$ and $y$ do not reach their stationary values, and the temporal fluctuations of $x$ and $y$ are $O(1)$ in the classical dynamics. 
Our conjecture discussed in Sec.~\ref{sec:statistics} implies that, in the equilibrium state within the TSS of the original quantum system, the fluctuations of the corresponding macroscopic intensive variables are also $O(1)$ and do not vanish in the limit of $N\rightarrow\infty$.
Anomalously large fluctuations in an equilibrium state within the TSS are generic feature of fully connected models.

The important role played by the TSS should be emphasized.
In the TSS, a fully connected Ising ferromagnet is reduced to a semiclassical system with a few-body degrees of freedom.
It is unclear to what extent our result survives beyond the TSS, which is an open problem.
In particular, it is very important but very difficult open problem to understand the relation between the long-time average along an individual classical trajectory and the quantum average in an individual energy eigenstate for genuinely many-body systems beyond the analysis in the TSS~\cite{Castiglione1996}.

In experiment, long-range Ising models with the pair interactions decaying as $1/r^{\alpha}$ with $0\leq\alpha<d$, where $r$ is the distance and $d$ is the spatial dimension, have been realized in trapped ions~\cite{Porras2004,Britton2012,Islam2013,Richerme2014}.
Our result corresponds to the case of $\alpha=0$, but it is expected that the system with $0<\alpha<d$ would have some common feature with the system with $\alpha=0$ like in equilibrium~\cite{Mori2011_instability,Mori2012_equilibrium} although there is no permutation symmetry and therefore the TSS is no longer invariant in the case of $0<\alpha<d$.
Thus, studying the case of $0<\alpha<d$ is an experimentally relevant interesting problem~\footnote{T. Mori, in preparation}, which might help us what happens when we go beyond the TSS. 

We have only considered local observables, but it is also important to consider purely quantum non-local quantities like the entanglement entropy, which is a purely quantum object and whose dynamics has been studied for the LMG model~\cite{Vidal2004}.

\begin{acknowledgments}
The author thanks the anonymous referees for helpful comments.
This work was financially supported by JSPS KAKENHI Grant No.~15K17718.
\end{acknowledgments}

\appendix

\section{Poincare section in the spin-1 fully connected Ising ferromagnet}
\label{sec:Poincare}

\begin{figure*}[tb]
 \begin{center}
\begin{tabular}{ccc}
\includegraphics[width=50mm]{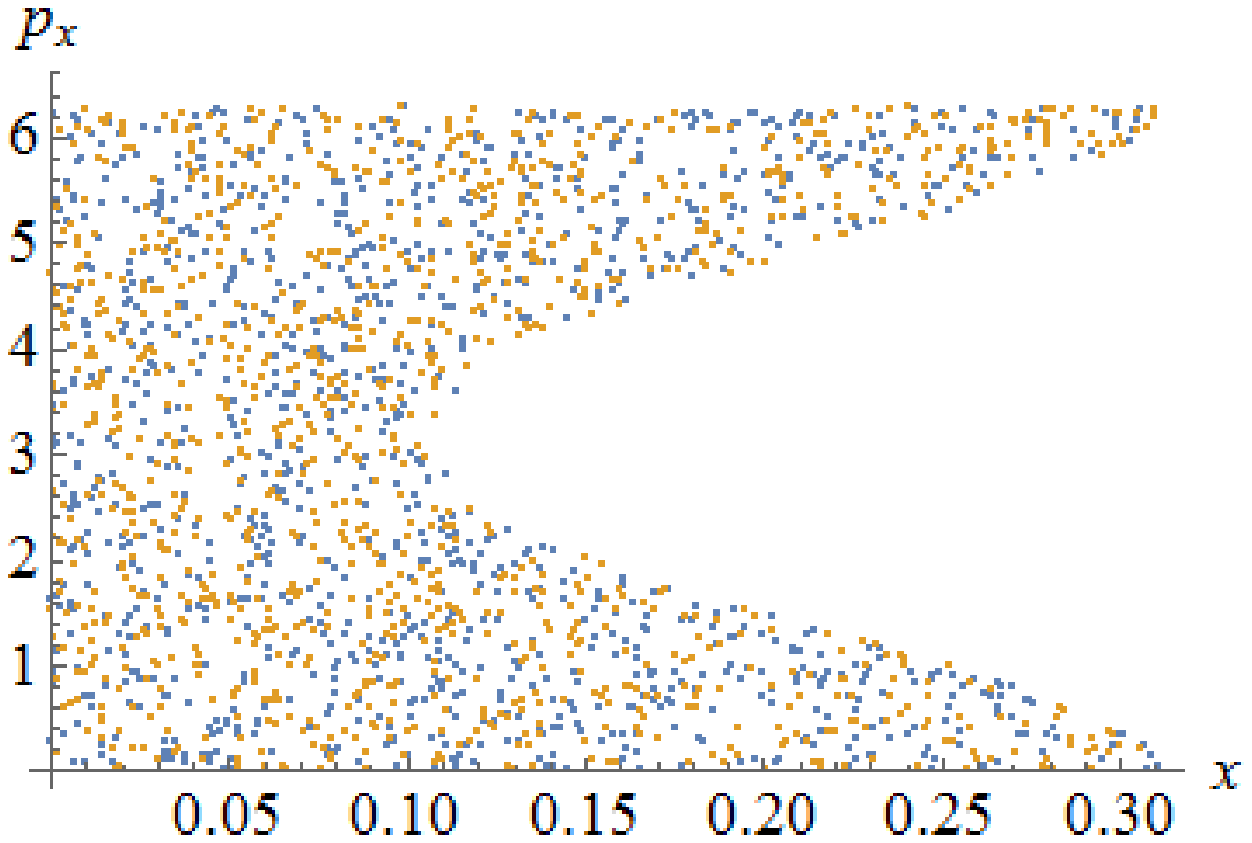}&
\includegraphics[width=50mm]{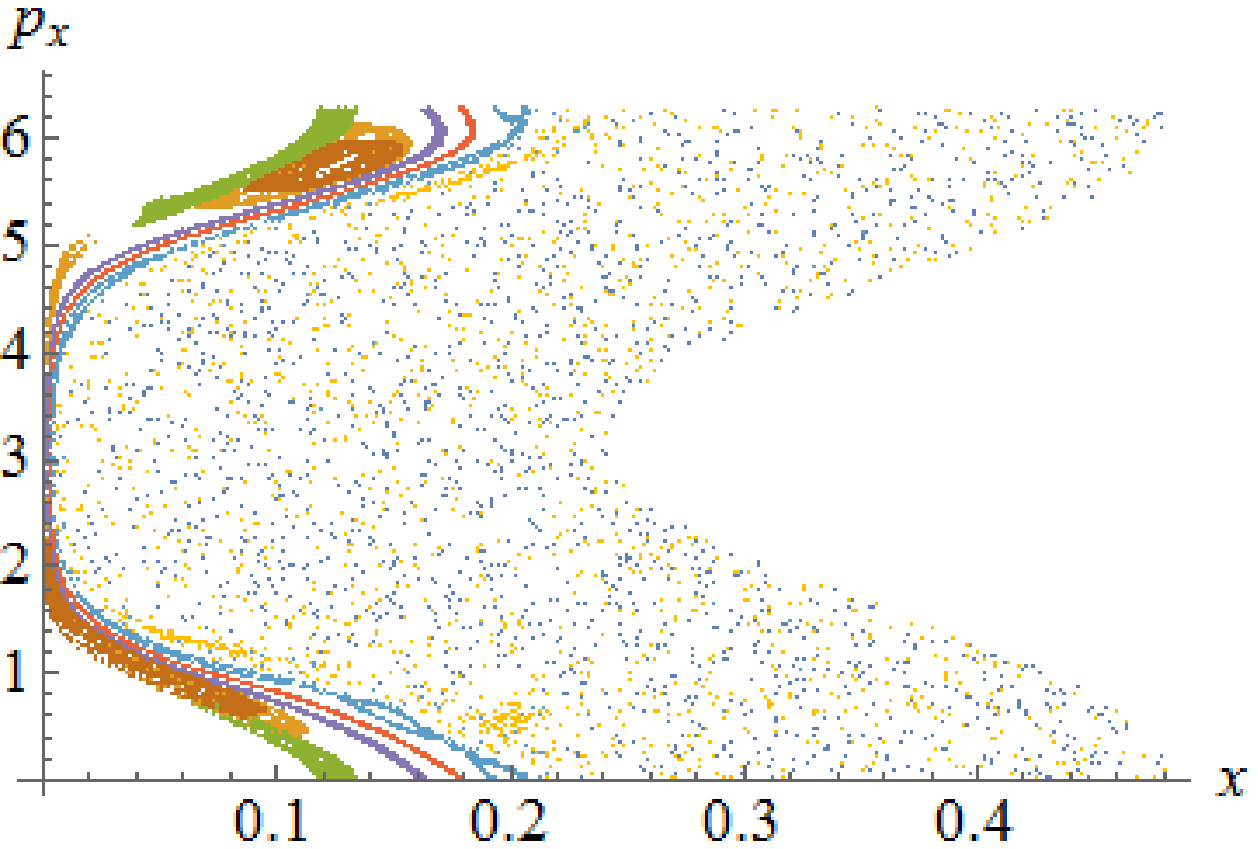}&
\includegraphics[width=50mm]{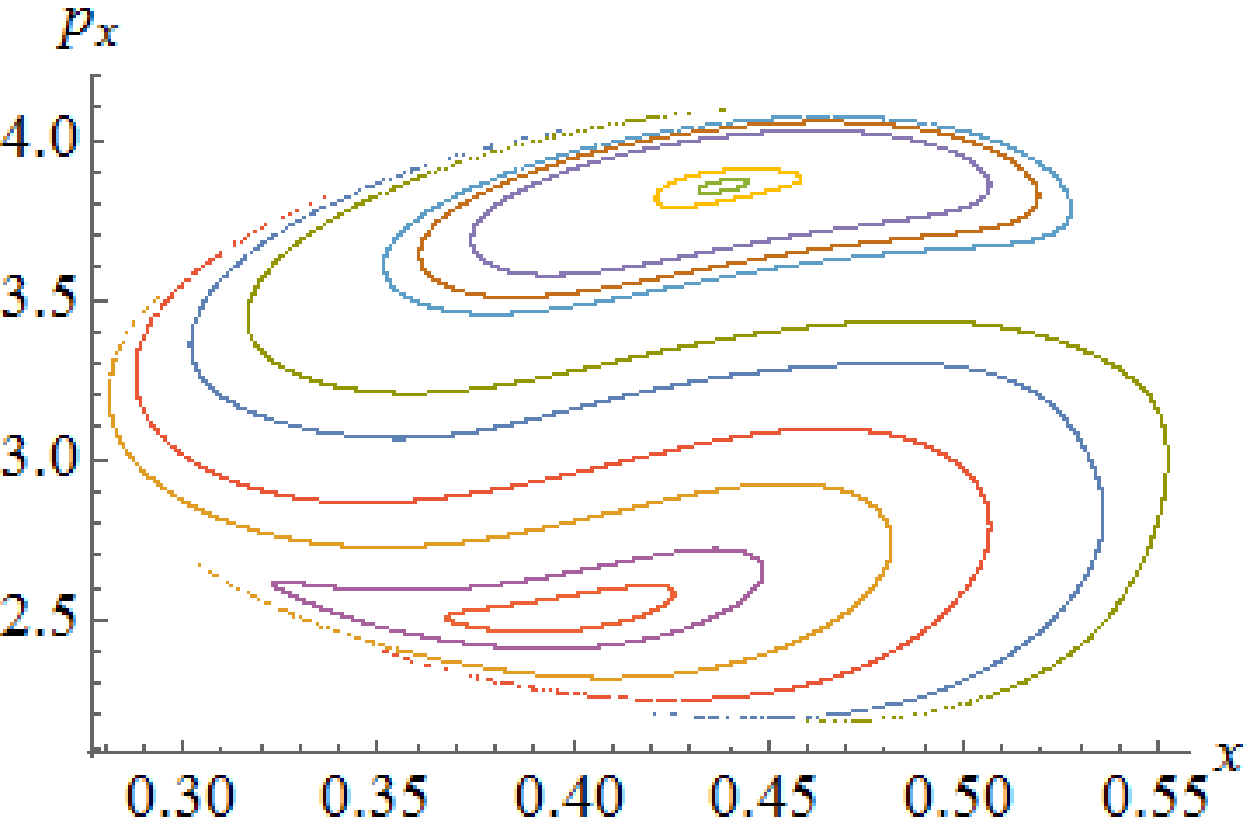}
\end{tabular}
\end{center}
 \caption{Poincare section in the classical dynamics of the spin-1 fully connected Ising ferromagnet.
Different colors imply different initial states in each figure.
The energy densities are 0.1 (left), 0.22 (middle), and 0.45 (right).
The parameters are identical to those in the main text, $h_x=0.2$, $h_z=0.01$, and $D=0.4$.}
 \label{fig:Poincare}
\end{figure*}

In contrast to the spin-1/2 case, the classical dynamics in the spin-1 case is complicated because there are two degrees of freedom ($x$ and $y$) and the equal-energy surface is three dimensional, in which the dynamics can be regular or chaotic.
In order to help the understanding of the classical dynamics in the spin-1 case, the Poincare section is presented in Fig.~\ref{fig:Poincare}, in which the energy density is set as $\varepsilon=0.1$ (left), $\varepsilon=0.22$ (middle), and $\varepsilon=0.45$ (right).
Different colors imply different initial states.
The Poincare surface is defined as the surface satisfying $\sin p_y\geq 0$ and $y=0.4$.
In the case of $\varepsilon=0.1$, the Poincare section implies that the dynamics is fully chaotic.
In the case of $\varepsilon=0.22$, in which the ergodicity is strongly violated (see Fig.~\ref{fig:classical-quantum}), the chaotic region and the regular region coexist.
In the case of $\varepsilon=0.45$, the classical dynamics is regular.

\section{The result in the Dicke model}
\label{sec:Dicke}

\begin{figure*}[tb]
 \begin{center}
  \includegraphics[width=140mm]{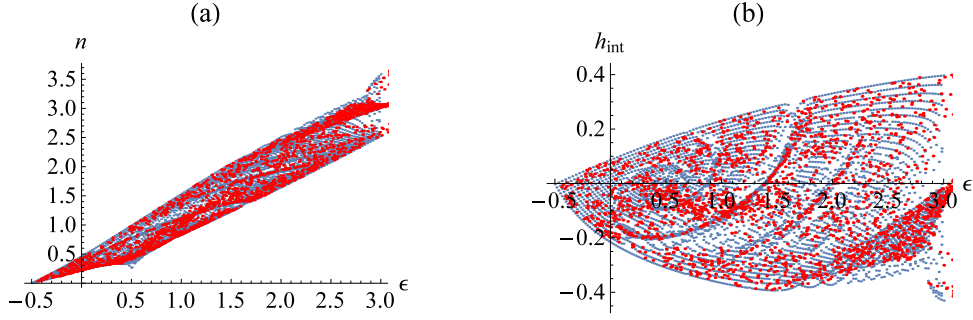}
\end{center}
 \caption{Long-time averages of (a) $n=a^{\dagger}a/N$ and (b) $h_{\mathrm{int}}$ in the classical dynamics starting from random initial states sampled uniformly from the phase space (red points) and expectation values of the same quantities in each quantum energy eigenstates for $N=40$ (blue points). The horizontal axis is the energy density $\varepsilon$.
The number of the samples of the random initial states in the classical dynamics is 2000. The classical dynamics is calculated up to $t=10000$ and the long-time averages are calculated within this time interval.}
 \label{fig:Dicke}
\end{figure*}

We consider the Dicke model, whose Hamiltonian is given by
\beq
H_D=\omega_pa^{\dagger}a+\omega_a\sum_{i=1}^NS_i^z-\frac{g}{\sqrt{N}}(a+a^{\dagger})\sum_{i=1}^NS_i^x,
\eeq
where $a$ and $a^{\dagger}$ are the annihilation and the creation operators of cavity photons, and $\bm{S}_i$ is the spin-1/2 operator of $i$th spin.
The collective interaction between cavity photons and $N$ spins is given by $g$.
Here we set $\omega_p=\omega_a=1$ and $g=0.6$.

In the TSS, the basis state $|x,y\>$ is characterized by the two variables $x$ and $y$, where
\beq
\frac{1}{N}a^{\dagger}a|x,y\>=x|x,y\>, \qquad
\frac{1}{N}\sum_{i=1}^NS_i^z|x,y\>=y|x,y\>.
\eeq
The Schr\"odinger equation for the wave function $\psi_t(x,y):=\<x,y|\psi(t)\>$, where $|\psi(t)\>$ obeys $i\d|\psi(t)\>/\d t=H_D|\psi(t)\>$, is obtained as, in the leading order in $N$,
\begin{align}
\frac{i}{N}\frac{\d}{\d t}\psi_t(x,y)=&\Bigg[\omega_px+\omega_ay
\nonumber \\
&\left.-2g\sqrt{x\left(\frac{1}{4}-y^2\right)}\cos p_x\cos p_y\right]\psi_t(x,y)
\nonumber \\
=:&\bar{H}_D\psi_t(x,y),
\end{align}
where $p_x=-(i/N)\d/\d x$ and $p_y=(-i/N)\d/\d y$ are the canonical momenta conjugate to $x$ and $y$, respectively.
It is noted that $1/N$ plays the role of the Planck constant $\hbar$.

\begin{figure}[tb]
\begin{center}
  \includegraphics[width=70mm]{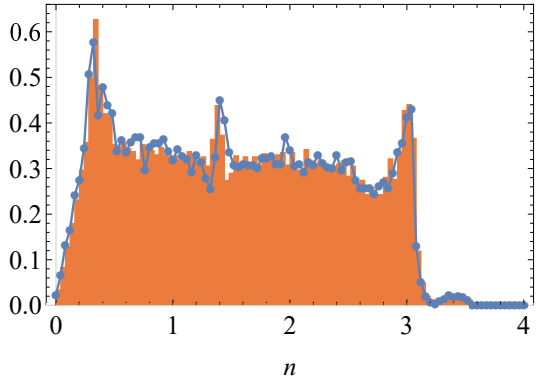}
\end{center}
 \caption{The distribution of the expectation values of $n=a^{\dagger}a/N$ in individual energy eigenstates for $N=60$ (shaded histogram) and the distribution of the long-time average of $x$ in the classical dynamics (points and line).}
 \label{fig:Dicke_n_dist}
\end{figure}

\begin{figure}
\begin{center}
  \includegraphics[width=70mm]{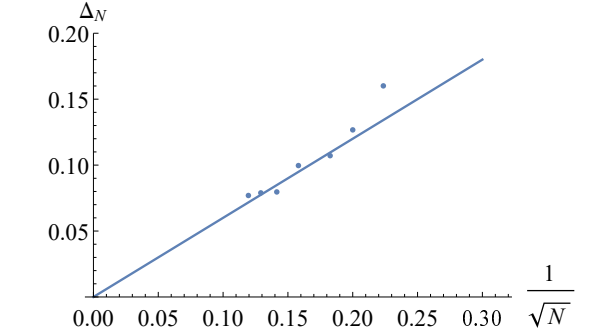}
\end{center}
 \caption{$\Delta_N$ as a function of $1/\sqrt{N}$ in the Dicke model. The smallest and the largest values of $N$ are 20 and 70, respectively.}
 \label{fig:Dicke_finite_size}
\end{figure}

In the limit of $N\rightarrow\infty$, the system becomes classical, and the corresponding classical equations of motion for $\{x(t),y(t),p_x(t),p_y(t)\}$ are given by the Hamilton equations under the classical Hamiltonian $\bar{H}_D$.
As in the spin-1 fully connected Ising ferromagnet, we calculate the distribution of quantum eigenstate expectation values of $n:=a^{\dagger}a/N$ and the interaction energy per spin,
\beq
h_{\mathrm{int}}:=\frac{1}{N}\left[-\frac{g}{\sqrt{N}}(a+a^{\dagger})\sum_{i=1}^NS_i^x\right],
\eeq
and the distribution of the long-time averages of the same quantities in the classical dynamics with random initial states sampled uniformly from the phase space.
In the calculation of quantum eigenstate expectation values, the number of cavity photons is truncated at $N_{\mathrm{max}}=6N$ in order to avoid the infinite dimension of the Hilbert space.
The result is presented in Fig.~\ref{fig:Dicke}, and one can see good agreement of the two distributions for both observables.
The probability distribution of $\<\phi_n|(a^{\dagger}a/N)|\phi_n\>$ (shaded histogram) and the probability distribution of the long-time average of $x(t)$ in the classical dynamics (points and line) are presented in Fig.~\ref{fig:Dicke_n_dist} for $N=60$, and the deviation $\Delta_N$ between them is shown in Fig.~\ref{fig:Dicke_finite_size} as a function of $1/\sqrt{N}$.
These results show that the conjecture discussed in Sec.~\ref{sec:statistics} also holds in the Dicke model.

\end{document}